\documentclass[particles,article,submit,pdftex,moreauthors]{mdpi} 

\firstpage{1} 
\pubvolume{1}
\issuenum{1}
\articlenumber{0}
\pubyear{2026}
\copyrightyear{2026}
\datereceived{ } 
\daterevised{ } 
\dateaccepted{ } 
\datepublished{ } 
\Title{Exploiting the latent space of deep AutoEncoders for the identification of signal pulses in noisy time-series}

\Author{Gioacchino Alex Anastasi$^{1,2,}$*\orcidA{}, Sebastiano Francesco Albergo$^{1,2,3}$\orcidB{}, Marzio De Napoli$^{1,2}$\orcidC{}, Noemi Pino$^{1,4}$\orcidD{}, Sebastiana Maria Puglia$^{1,2,3}$\orcidE{} and Alessia Rita Tricomi$^{1,2,3}$\orcidF{}}

\AuthorNames{Gioacchino Alex Anastasi, Sebastiana Maria Puglia, Marzio De Napoli, Alessia Rita Tricomi and Sebastiano Francesco Albergo}

\address{%
$^{1}$ \quad Dipartimento di Fisica e Astronomia "E. Majorana", Università di Catania, Via Santa Sofia 64, Catania, Italy\\
$^{2}$ \quad Istituto Nazionale di Fisica Nucleare (INFN), Sezione di Catania, Via Santa Sofia 64, Catania, Italy\\
$^{3}$ \quad 
Centro Siciliano di Fisica Nucleare e di Struttura della Materia (CSFNSM), Via Santa Sofia 64, Catania, Italy\\
$^{4}$ \quad 
Istituto Nazionale di Fisica Nucleare (INFN), Laboratori Nazionali del Sud, Via Santa Sofia 62, Catania, Italy
}

\corres{Corresponding author, gialex.anastasi@dfa.unict.it}

\abstract{
We propose a data-driven procedure, based on convolutional variational autoencoders, to identify the presence of signal pulses in long time series. The dataset consists of synthetic waveforms, each composed of non-Gaussian noise and a log-normal-shaped signal of variable intensity, with a length of 10,000 samples. The model heavily compresses the input waveforms, allowing a direct study of such a reduced representation. After training for 150 epochs on 7500 waveforms, a region in the latent space where the network encodes time-series presenting only background noise emerges, allowing, in turn, to tag as candidates for containing a signal those falling outside. 
When applied to a test dataset of freshly generated waveforms, 100\% of events with signal amplitudes well above the baseline noise are correctly labelled, and this fraction only decreases for amplitudes comparable with accidental noise pulses.
This approach was designed to fully exploit the measurements in dual-phase Liquid Argon Time Projection Chambers, as the one of the Recoil Directionality experiment, built in the context of the Darkside project.
}


\begin{document}


\section{Introduction}

In the era of big data and artificial intelligence, research at the forefront of fundamental physics is increasingly becoming a computing-intensive field. Modern and upcoming experimental facilities generate large data volumes and exhibit processing requirements that rival those of the world's leading technological enterprises. While this trend has historically driven the evolution of computing infrastructures within the high-energy physics community, comparable resource demands have now emerged across other domains, most notably gravitational wave interferometry and astroparticle physics.
Furthermore, recent research has shifted toward unsupervised or weakly supervised neural network frameworks in order to fully exploit the available data, with~successful implementations in the search for anomalous signatures in high-energy collider measurements~\cite{paper_ATLAS, paper_CMS} and in the mitigation of non-Gaussian noise transients in gravitational wave detectors~\cite{paper_GW,paper_GW2}. A
comprehensive repository of publications concerning machine learning applications in particle physics and neighbouring fields can be found in~\cite{ML_living}.

Exporting these data-driven paradigms to astroparticle physics is one of the core objectives of the DAIDREAM (``{DAta-driven IDentification of Rare Events in Astroparticle physics through Machine learning techniques'') use-case~\cite{paper_CRISMAC}, an~initiative within Work Package 3 (``Experimental Astroparticle Physics and Gravitational Waves'') of the Spoke 2 (``Fundamental Research \& Space Economy''), under~the umbrella of the \textit{Italian National Research Centre for High Performance Computing, Big Data and Quantum Computing} (ICSC). 
The primary goal of DAIDREAM is to implement unsupervised neural architectures which, after~being properly adapted to the required experimental setting, can maximize the discovery potential of current and next-generation rare-event~searches. 

A promising field of application of this approach is the detection of faint signals produced by particles traversing a Liquid Argon Time Projection Chamber (LAr TPC). Such apparatuses offer excellent calorimetric capabilities and high-resolution position reconstruction~\cite{paper_Rubbia}, and~thus has been chosen for many accelerator-based neutrino experiments, such as ArgoNeuT~\cite{paper_ArgoNeuT}, the~Short-Baseline Neutrino Oscillation program at Fermilab~\cite{paper_SNO1, paper_SNO2} and DUNE~\cite{paper_DUNE}. 
The same technology is also employed in direct Dark Matter (DM) searches, where candidate events are identified via nuclear (and, in~specific scenarios, electronic) recoils in argon.
The latter research, in~particular in the context of the DarkSide program, is the primary focus of this paper. The~goal is to overcome the limitations of traditional pulse-finding algorithms, which rely on predefined thresholds, by integrating instead methodologies rooted in advanced deep learning into the analysis~pipelines.

In this work, developed within the DAIDREAM framework, we introduce a methodology aimed at identifying faint signals in noble-liquid TPCs by leveraging unsupervised representation learning, rather than supervised procedures as in~\cite{paper_proof_of_concept,paper_ROI,paper_ml_xenon,paper_Chung}.
In particular, we focus on the data-driven analysis of noisy time series acquired by photodetectors observing a dual-phase (liquid and vapour) argon TPC.
To this end, we designed a custom Variational Convolutional Autoencoder (VCAE) architecture that encodes generic raw digitized waveforms into a low-dimensional latent space, effectively capturing the fundamental information from each time series. 
While this class of deep neural networks has been widely used for anomaly detection and new physics searches at colliders~\cite{paper_LHC_VAE,paper_taxonomic}, the~novelty of our approach lies in constructing a signal-selection criterion that directly exploits the learned compressed representation, which is rarely studied in the field~\cite{paper_LHC_VAE2,paper_LZ_VAE}.

In Section~\ref{sec_exp}, the~scientific and experimental context is described, together with the design of a dataset of synthetic waveforms that mimics the related noise conditions and pulse topologies. Section~\ref{sec_VAE} deals with the deep neural network architecture, training and optimization. In~Section~\ref{sec_latent}, we demonstrate that the encoded representation of waveforms containing only a negligible signal naturally clusters into a dense, restricted region of the latent space. Based on this emerging property, we study the Euclidean distance from the centroid of the encoded representations of noise-only time-series. Such a simple metric allows for an event-by-event identification of low-amplitude signal pulses, with~100\% efficiency above a specific threshold in the signal amplitude. The~obtained results are discussed in Section~\ref{sec_results}, and~a summary of the proposed representation learning framework, with~an outlook about the applications to real measurements, is reported in Section~\ref{sec_conclusions}.

\section{Experimental Context and Synthetic Waveform~Generation\label{sec_exp}}

\textls[-25]{While the existence of dark matter is supported by several observational evidences~\cite{paper_Rubin, paper_Clowe, paper_Planck}}, the~nature of its component(s) remains unknown.
Searches for dark matter candidates can be roughly classified in indirect detection in astrophysical environments, production at accelerators and colliders, and~direct detection in experimental apparatuses designed to observe the scatter or interaction of DM candidates with nuclei or electrons in the detector material.
Comprehensive review of the most compelling (although elusive) particle candidates beyond the Standard Model can be found in~\cite{PDG_dark_matter, paper_particle_DM}, including axions and axion-like particles, sterile neutrinos, dark photons, and~Weakly Interacting Massive Particles (WIMPs). Other models predict composite DM candidates, like axion quark nuggets~\cite{paper_AQN} and ultra-heavy dark matter~\cite{paper_UHDM}, or~``rich'' dark sectors where DM particles interact through new mediators~\cite{paper_rich_dark_sector}. 

In the context of direct searches, the~detection of WIMPs with a mass in the range from a few GeV/c$^2$ up to tens of TeV/c$^2$ has been the primary focus for decades, as these particles are theoretically favoured~\cite{paper_WIMPs_paradigm} although only interacting with standard matter via weak interactions (in addition to gravity).
The most mature technologies employed in this field are Liquid Xenon and Liquid Argon TPCs: both elements are ideal targets for rare event detection, being scintillators highly transparent to their own scintillation light; at the same time, the~TPC apparatus provides the event position, allowing the definition of a low-background inner fiducial volume.
Regarding Liquid Xenon TPCs, the~LZ experiment~\cite{paper_LZ} presently holds the most stringent constraints for WIMP masses above 9~GeV/c$^2$, with~other two multi-tonne apparatuses---XENONnT~\cite{paper_XENONnT} and PandaX-4T~\cite{paper_PandaX}---operating with a similar sensitivity.
Instead, the~DEAP-3600~\cite{paper_DEAP} and DarkSide-50 (DS-50) \cite{paper_DS50} experiments demonstrated the feasibility of multi-tonne-scale LAr detectors, in~particular using Underground Argon (UAr) \cite{paper_DS50_UAr}, heavily depleted in the radioactive isotopes. More than 400 scientists, from~over 60 institutions, have constituted the Global Argon Dark Matter Collaboration (GADMC) to pursue the search for DM using a UAr target with the DarkSide-20k (DS-20k) \cite{paper_Ds20k} and Argo experiments.
More information and references on this topic can be found in~\cite{snowmass_report1, snowmass_report2}.

In this work, we focus on dual-phase (liquid and vapour) LAr TPCs, where a scattering event is typically identified by observing two distinct signals~\cite{paper_S1_S2} via photosensors: a prompt scintillation light pulse ($S1$) from the liquid phase, and~a delayed electroluminescence signal ($S2$) from ionization electrons drifted and extracted into the gas phase, as~represented in Figure~\ref{f1}a. This technology has been employed in the DS-50 experiment~\cite{paper_DS50_TPC}, and~it is the main component of the future DS-20k apparatus~\cite{paper_Ds20k}.

However, considering the lack of DM detection, recent theoretical and experimental interest also includes lighter candidates, with~masses as low as $1\ \text{GeV}/\text{c}^2$ \cite{paper_Ds50_lowmass, paper_Ds20k_lowmass}. 
Exploring this range remains an experimental challenge, even with the mature technology of dual-phase LAr TPCs, primarily because very small energy deposits are predicted in this scenario: for low-mass WIMPs, the~nuclear recoil energy would be limited to a few $\text{keV}$. At~such regime, the~prompt scintillation signal ($S1$) becomes exceedingly weak, forcing reliance almost exclusively on the delayed electroluminescence signal ($S2$).

In the DS-50 experiment, $S1$ and $S2$ signals were identified by applying a pulse-finding algorithm with a fixed threshold on an overall waveform, obtained by summing all the calibrated measurements from each photomultiplier tube looking at the TPC. More in detail, a~baseline is determined for each channel with a moving average algorithm, and~subtracted from the raw waveforms; then all channels are calibrated, scaling the baseline-subtracted waveforms by the corresponding single photoelectron (SPE) response, and~added together. The~summed waveform is used for pulse finding, employing a coarse-grained search to discern pulses and a fine-grained search to identify pulse start times. The~complete procedure is reported in~\cite{paper_DS50_TPC}.
In Figure~\ref{f1}b, the~spectrum of the $S2$ signals is reported in terms of detected ionization electrons $N_\texttt{e}$, defined as the integrated number of $S2$ photoelectrons (PE) divided by the $S2$ yield (that is, the mean number of photoelectrons per ionization electron extracted in the gas pocket, equal to 23 $\pm$ 1~PE/e$^-$). 
The search for low-mass WIMPs required several quality and selection cuts, described in detail in~\cite{paper_Ds50_lowmass}, which affect the lower energy part of the dataset.
The range of interest goes from 3~to 170~$N_\texttt{e}$, corresponding to [0.04,~21]~keV for an electron recoil (ER), or~[0.4,~393]~keV for a nuclear recoil (NR).  
In any case, the~hardware trigger efficiency and software pulse-finder efficiency are reported~\cite{paper_Ds50_lowmass_old} to be essentially 100\% for $S2$ signals larger than 30 PE, which correspond to about 1.5~e$^-$, well below the minimum of the reported range of~interest.

\vspace{-6pt}
\begin{figure}[H]
\begin{adjustwidth}{-\extralength}{0cm}
\centering

\subfloat[\centering]{\label{fig_TPC}\includegraphics[width=6.5cm]{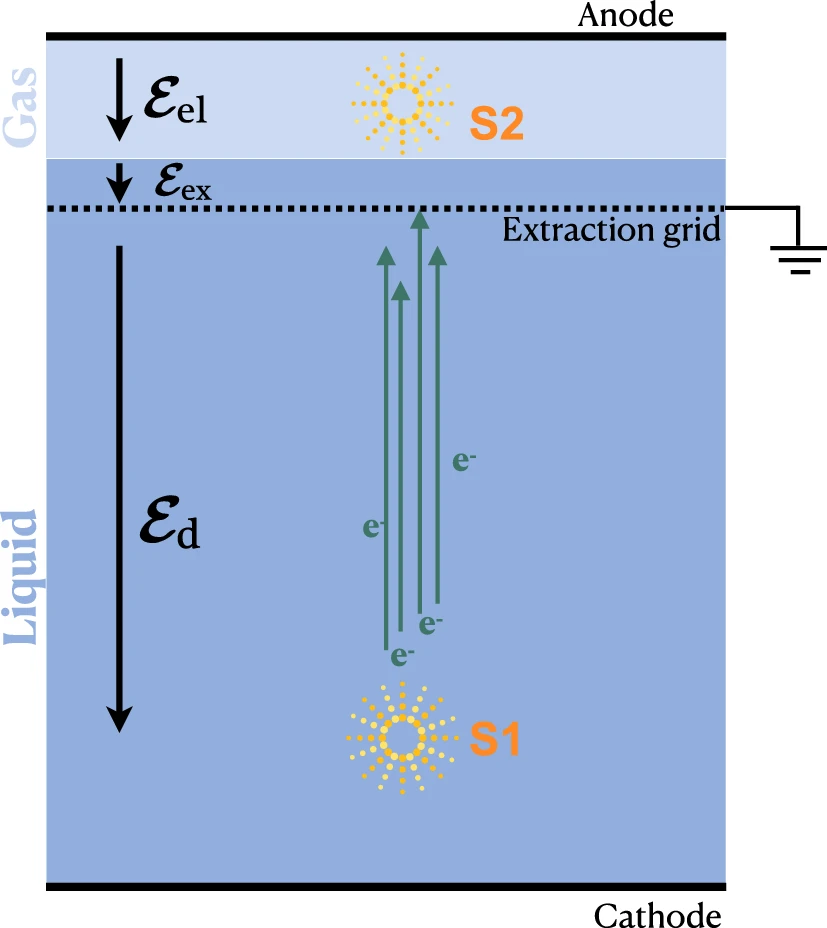}}
\subfloat[\centering]{\label{fig_lowmass}\includegraphics[width=11.0cm]{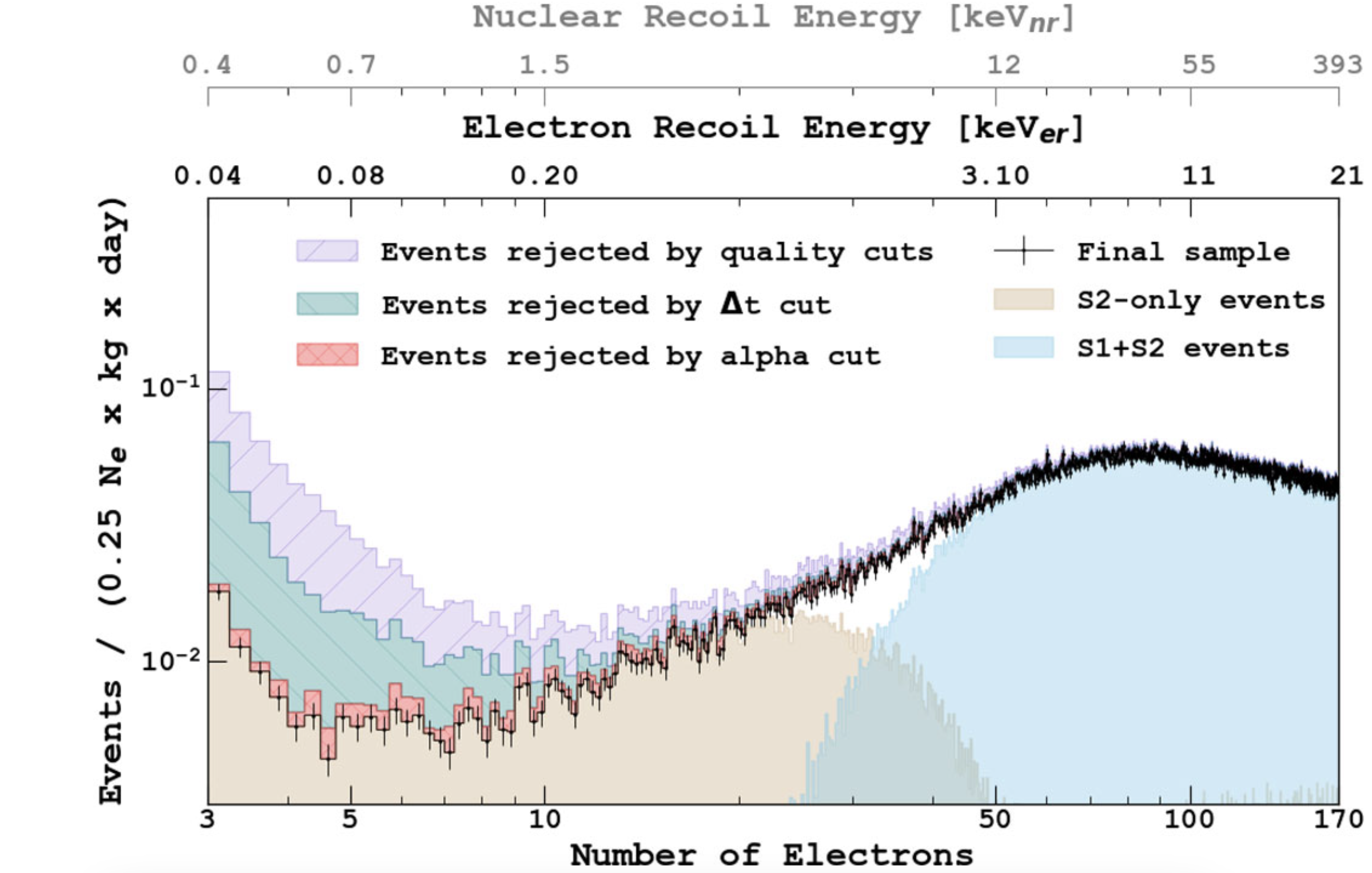}}

\end{adjustwidth}

\caption{(\textbf{a}) Sketch from~\cite{paper_ReD_TPC}, representing the detection of a recoil event in a double-phase LAr TPC, where both the scintillation signal ($S1$) and the delayed electroluminescence signal ($S2$) are present. To~generate the $S2$ pulse, three different electric fields are maintained inside the TPC: the drift field ($\mathcal{E}_d$) to drift ionisation electrons towards the extraction grid; the extraction field ($\mathcal{E}_{ex}$) to extract them into the gas phase; the electroluminescence field ($\mathcal{E}_{el}$) to produce the electroluminescence signal in the gas.
(\textbf{b}) Spectra of the number of detected electrons $N_e$ (see text for the definition) at different steps of the data selection for the Darkside-50 low-mass WIMPs analysis described in~\cite{paper_Ds50_lowmass}.}\label{f1}
\end{figure} 

In Figure~\ref{f1}b, it can also be noticed that, for~NR energies below $\sim$10~keV$_{nr}$, the~final dataset is mostly composed of $S2$-only events.
Such low-threshold analyses depend critically on a correct evaluation of the ionization yield $Q_y$ of nuclear recoils in argon, defined as the number of ionization electrons produced per unit of recoil energy.
The described low-mass WIMP search in DS-50 employed an ad~hoc phenomenological model, extrapolating the $Q_y$ values down to $\sim$0.5~keV~\cite{paper_Ds50_calibration}. 
Instead, a~direct measurement of $Q_y$ in the same energy range is fundamental, also in view of the DS-20k~experiment.

To systematically investigate the region below $5\text{ keV}$, previously not explored in the literature, a~dedicated measurement campaign has been accomplished within the \textit{Recoil Directionality} (ReD) experiment~\cite{paper_ReD_TPC, paper_ReD_dir} at the INFN Catania Section. 
The ReD apparatus features a miniaturized version of the DS-20k LAr TPC, with~an active volume of \mbox{$5 \times 5 \times 6\text{ cm}^3$}, designed to detect the elastic scattering of neutrons produced by a Californium-252 source.
To tag fission events and reconstruct the neutron kinematics, the~ReD setup presents: (i)~Barium Fluoride ($\text{BaF}_2$) detectors in proximity to the source identify $\gamma$-rays accompanying neutron emission; (ii) a spectrometer composed by 18 scintillators about 100~cm beyond the TPC, to~detect the scattered neutrons and their position.
This setup, together with the choice of the source, was designed to measure NRs releasing 1--10~keV in the TPC while precisely reconstructing the kinematics of the corresponding~neutrons.

The signals inside the TPC are acquired by two tiles of 24 customised cryogenic silicon photomultipliers (SiPMs) \cite{paper_SiPM}, one tile placed at the top and one at the bottom. The~SiPM channels are sampled at $500\text{ MHz}$ and a full acquisition window encompasses 40,000 samples, for~a total duration of 80 \textmu s. 
In the offline analysis, the~individual raw waveforms are pre-processed by numerically deconvolving the time response to the single photoelectron separately for each SiPM, calculated from laser calibration runs.
The deconvolved waveforms are summed and a pulse-finding algorithm is then applied to the overall waveform to identify $S1$ and $S2$ pulses.
The pulse finder acceptance, evaluated by generating Monte Carlo deconvolved waveforms, is about 60\% for $S2$ signals of 30 PE (corresponding to approximately 1.5 ionization electrons), reaching 100\% above 60 PE~\cite{paper_ReD_results}.

The final analysis provided a direct and model-independent measurement of the ionization yield in argon for nuclear recoils between 2 and 10 keV, in~good agreement with previous measurements at 6.7~keV while showing an increase at the lowest energies~\cite{paper_ReD_results}.
Improving our understanding of the LAr ionization response at low recoil energies translates into stronger WIMP exclusion limits and enhanced projected sensitivity for the DS-20k project~\cite{paper_ReD_results2}. For~this reason, an~upgrade of the ReD apparatus (called ReD+) is planned to further pursue this~investigation.

In the context of small-scale experiments, distinguishing a genuine $S2$ pulse from stochastic baseline fluctuations, electronic noise, or~spurious single-electron extractions still represents a challenge, given the low signal amplitude at nuclear recoil energies below 1~keV.
Such a task can be effectively tackled using methodologies rooted in machine learning. However, to~systematically evaluate the performance and limits of a deep neural network framework under controlled conditions, a~synthetic dataset mimicking the $S2$-only waveforms collected by ReD was developed, as~described in the next~paragraph.

\subsection*{Toy Model for Low-Energy Pulse and Noise~Generation \label{subsec_toy}}

In order to characterize $S2$-only signal pulses emitted in the gas pocket, the~first step is to produce synthetic waveforms resembling the measurements of the photosensors in the upper part of the TPC.
Furthermore, to~ensure the applicability of our deep learning methodology across different experimental set-ups, we decided to generate an overall waveform that approximately reproduces the average of the calibrated time-dependent signals (a standard and generally useful pre-processing output, considering, for instance, the DS-50 analysis).

Each synthetic waveform $W(t)$ is generated using the \texttt{scipy.stats} library~\cite{url_SciPyStats} and is modelled as the superposition of a non-Gaussian stochastic background $B(t)$ and a single log-normal signal pulse $S(t)$:
\begin{linenomath}
\begin{equation}
W(t) = B(t) + S(t)
\label{eq_Wt}
\end{equation}
\end{linenomath}
The length is fixed to 10,000 time-bins to optimize the computational efficiency and reduce the memory footprint during the deep learning training phases. 
This would correspond to downsampling the ReD measured waveforms by a factor of four (by calculating the mean every 4 consecutive time-bins), which anyway preserves the integrity of the $S2$ pulse while drastically reducing the downstream computational load.

The $S2$-like pulse is modelled using a log-normal probability density function, which reproduces the typical fast rise (order of a few nanoseconds) and the slower tail (order of \mbox{1 \textmu s}) of electroluminescence signals:
\begin{linenomath}
\begin{equation}
\vspace{-0.1cm}
S(t) = A_{S} \cdot f_{\text{log-normal}}(t; \mu_{S}, \sigma_{S}, T_{S})
\label{eq_signal}
\end{equation}
\end{linenomath}    
For each instance, the~temporal position of the pulse $T_{S}$ is drawn uniformly, excluding the first $150$ and the last $150$ time-bins; also the log-normal location and shape parameters $\mu_{S}$, $\sigma_{S}$ are sampled uniformly from narrow intervals: 
\begin{linenomath}
\vspace{-0.1cm}
\begin{equation}
\mu_{S} \sim \mathcal{U}(6.65, 6.75) \qquad \sigma_{S} \sim \mathcal{U}(0.35, 0.45)
\vspace{-0.1cm}
\end{equation}
\end{linenomath}
These values have been inferred from the ReD data, more precisely by studying the pulse shape in the average of the calibrated waveforms acquired by the top tile SiPMs in nuclear recoil events with energies in the 1--10~keV~range.

Finally, the~signal is rescaled by a factor $A_{S}$, sampled independently for each profile as:
\begin{linenomath}
\begin{equation}
\log_{10}(A_{S}) \sim \mathcal{U}\left(0,\ \log_{10}\left\{\max_t \left[f_{\text{log-normal}}(t; \mu_{S}, \sigma_{S}, T_{S})\right]^{-1}\right\}\right)
\label{eq_signal_As}
\end{equation}
\end{linenomath}
a formulation which mathematically produces a maximum signal amplitude (peak) spanning from 1 down to $\sim$0.001, given the ranges chosen for $\mu_{S}$ and $\sigma_{S}$, thus allowing efficient exploration of the model's sensitivity even at the single-photoelectron~regime.

The background is instead constructed by randomly summing and subtracting 100~small, independent log-normal components:
\begin{linenomath}
\begin{equation}
B(t) = \sum_{i=1}^{100} (-1)^{i\bmod2} \cdot A_{noise} \cdot f_{\text{log-normal}}(t; \mu_{noise}, \sigma_{noise}, T_{noise}^{i})
\label{eq_Bt}
\end{equation}
\end{linenomath}
where $\bmod$ denotes the modulus operation, $A_{noise}=0.5$, $\mu_{noise}=6$, $\sigma_{noise}=0.2$ are fixed parameters defining the log-normal shape, and~the time position $T_{noise}^{i}$ of each noise contribution is distributed uniformly across the entire 10,000-bin interval. This choice successfully produces a fluctuating background where spurious noise peaks can often match or exceed the amplitude of small signal~pulses.

In Figure~\ref{fig_synthetic_wf}, some of the resulting time-series are shown, together with their two components.
Since the amplitudes are sampled uniformly (in the logarithm) down to very low values, a~large fraction of the injected pulses are non-zero (as $A_{S} > 0$) yet effectively submerged within the baseline noise, making them visually and statistically almost indistinguishable from pure background events. 
This choice is intentional: by populating the low-amplitude regime, the~network is trained to reconstruct waveforms precisely in the noise-dominated region, where signal identification is most~challenging.

\vspace{-9pt}
\begin{figure}[H]
\begin{adjustwidth}{-\extralength}{0cm}
\centering 
\includegraphics[width=0.42\textwidth]{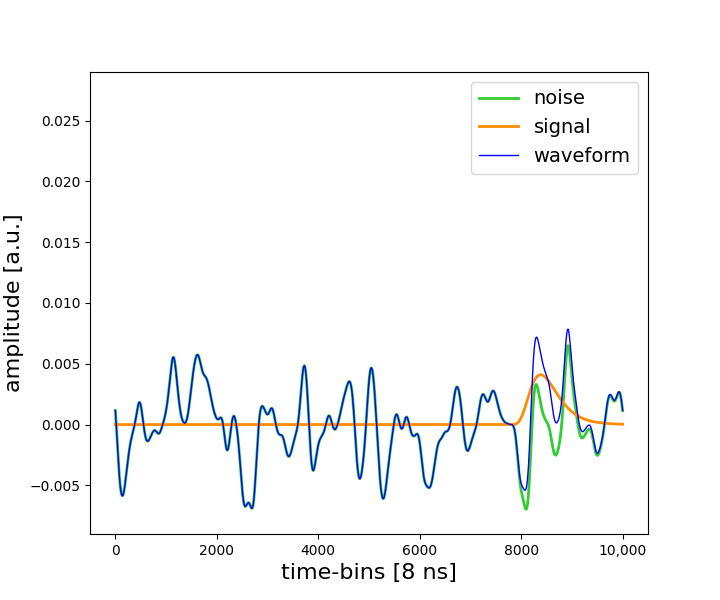}
\includegraphics[width=0.42\textwidth]{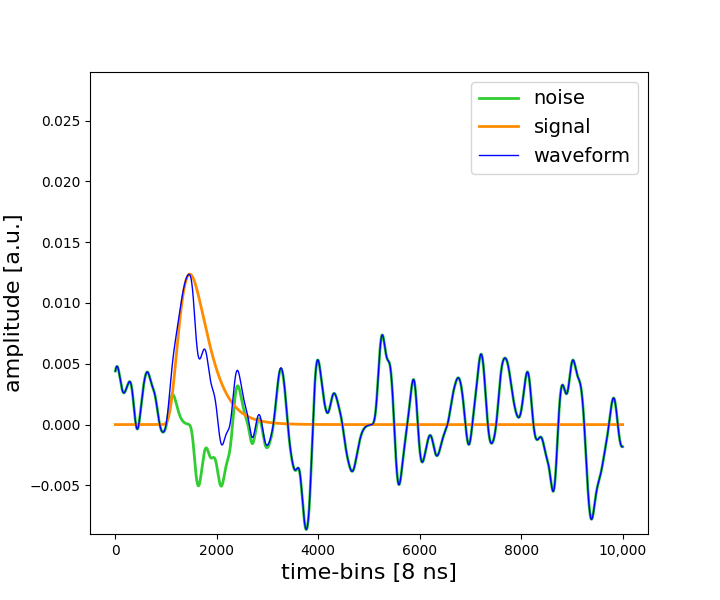}
\includegraphics[width=0.42\textwidth]{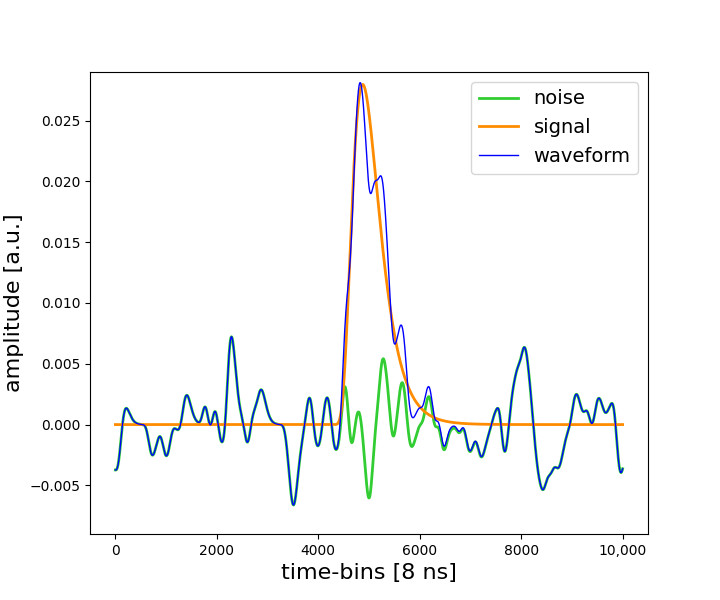}
\end{adjustwidth}
\caption{\label{fig_synthetic_wf} Examples of a synthetic waveform: in green the non-Gaussian noise $B(t)$, in~orange the log-normal shaped signal pulse $S(t)$, in~blue the final time-series $W(t)$.}
\end{figure}
\unskip

\section{Variational Convolutional~Autoencoder\label{sec_VAE}}

With the aim of developing a non-supervised deep learning application capable of identifying low-amplitude ionization signals produced by low-energy nuclear recoils within a LAr TPC, an~autoencoder (AE) architecture~\cite{book_Goodfellow_ch14} has been chosen. This class of networks is widely recognized in the literature for its versatility in key physics tasks such as denoising, anomaly detection, and~feature extraction~\cite{book_Bishop}.

In its standard formulation (see Figure~\ref{fig_autoencoder}), an~AE is a feedforward neural network divided into two symmetric components: an encoder and a decoder. The~encoder compresses the high-dimensional input data $\textbf{x}$ (in this context, a~waveform) into a lower-dimensionality vector $\textbf{z}$, belonging to the so-called latent space, bottleneck, or~coding layer. Vice versa, the~decoder attempts to reconstruct the original input from this compressed latent representation, delivering an output $\hat{\textbf{x}}$.
This paradigm forces the network to extract the fundamental but implicit characteristics of the dataset within the restricted information bottleneck.
The ``goodness'' of the predictions is then evaluated by directly comparing the input $\textbf{x}$ with the reconstruction $\hat{\textbf{x}}$, where the Mean Squared Error (MSE) is typically chosen for continuous time-series data:
\begin{linenomath}
\begin{equation}
\mathcal{L}_{\text{rec}} = \frac{1}{N} \sum_{i=1}^{N} \sum_{j=1}^{M} (x_j^{(i)} - \hat{x}_j^{(i)})^2
\label{eq_MSE}
\end{equation}
\end{linenomath}
where $N$ is the overall number of waveforms in the dataset, and~$M$ is the number of time-bins. 
The weights and biases of both the encoder and decoder are updated via gradient descent and backpropagation to minimize the MSE loss during the training phase.
As the input data themselves are employed to supervise the learning process, an~AE is usually referred to as a self-supervised~network.

\vspace{-6pt}
\begin{figure}[H]
\includegraphics[width=0.8\textwidth]{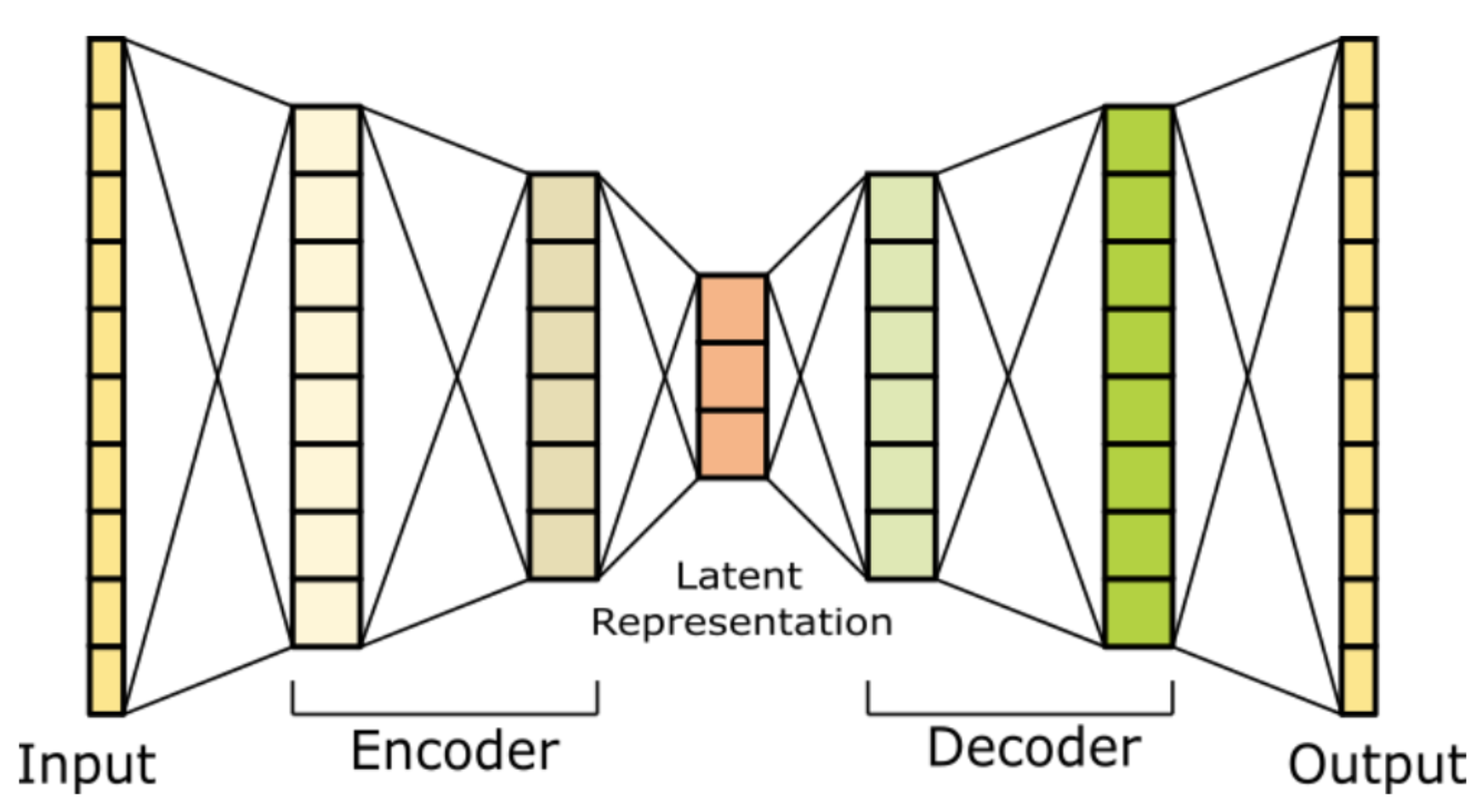}
\caption{\label{fig_autoencoder} A diagram illustrating the architecture of a standard deep~autoencoder.}
\end{figure}

In short, standard autoencoders map an input to a discrete, single point in the latent space. While this is effective for reconstruction/denoising tasks, it usually results in a discontinuous distribution of the encoded representations. Such irregularity hinders analyses where the topology and density of the latent space must be exploited for physical event classification, characteristic of representation learning tasks~\cite{book_Goodfellow_ch15}.

To overcome this structural limitation, we then implemented a Variational Autoencoder (VAE) \cite{paper_VAE,paper_VAE2} network.
A VAE, instead of compressing the input deterministically into a vector, aims to model the underlying probability distributions of the data. Given this paradigm, each input data point is mapped into a probability density function (p.d.f.) by having the encoder extract the p.d.f.~parameters. In~our case, we assumed Gaussian distributions, so that the encoder outputs a vector of means $\boldsymbol{\mu}$ and a vector of variances $\boldsymbol{\sigma^2}$.
The latent representation $\textbf{z}$ is then sampled from these distributions, by~extracting a value along each latent space dimension $k$ as:
\begin{linenomath}
\begin{equation}
z_k \sim \mathcal{N}(\mu_k(\textbf{x}), \sigma_k(\textbf{x}))
\end{equation}
\end{linenomath}
To enable backpropagation through this stochastic bottleneck, we employed the reparameterization trick~\cite{paper_VAE}, implementing a \texttt{Sampling} layer which defines $z_k = \mu_k + \sigma_k \cdot \epsilon$, where $\epsilon$ is an auxiliary variable drawn from a standard normal distribution $\epsilon \sim \mathcal{N}(0, 1)$. 

This probabilistic structure forces the latent space to be continuous and regularized, ensuring that nearby points correspond to physically similar events.
The features of the resulting latent representation can then be analysed to establish a selection criterion capable of labelling waveforms with pulses, as~described in Section~\ref{sec_latent}.

\subsection{Network Architecture and Loss~Function\label{subsec_arch}}

In this work, a~1D Variational Convolutional Autoencoder (VCAE) has been implemented via the Keras API~\cite{url_Keras} with TensorFlow backend~\cite{url_TS}.
\begin{itemize}
    \item \textbf{Encoder}: the input layer receives a (single-channel) vector. The~feature extraction is composed of three blocks, each comprising a 1D Convolutional layer (\texttt{Conv1D}) with Rectified Linear Unit (\texttt{ReLU}) activation function, followed by a \texttt{AveragePooling1D} layer. The~three convolutional layers are configured with 4, 8 and 16 filters, respectively, a~fixed kernel size of 32 and a stride of 4; the pooling is instead performed every 2~time-bins. Therefore, each block reduces the waveform length by a factor of 8 while increasing the channels.
    \item \textbf{Bottleneck}: after the third convolutional block, the~resulting 16 feature maps, each 1/64th of the initial waveform length, are flattened (\texttt{Flattening} layer) and then transformed into the latent mean vector $\boldsymbol{\mu}$ and the logarithm of the latent variance vector $\boldsymbol{\sigma^2}$ using two \texttt{Dense} layers. Finally, the~latent vector $\textbf{z}$ is extracted using a \texttt{{Sampling}} layer that implements the reparameterization trick.
    \item \textbf{Decoder}: the decoder strictly mirrors the encoder, beginning with a \texttt{Dense} layer that expands the latent vector $\textbf{z}$, followed by a \texttt{Reshape} layer which recovers the 16 compressed feature maps structure. Then three consecutive blocks of \texttt{UpSampling1D} and Transposed Convolutional (\texttt{Conv1DTranspose}) layers, with~8, 4 and 1 filters, respectively, perform the reconstruction of the initial waveform.
\end{itemize}
The bottleneck dimension is fixed to 2, as~early implementations using a single latent parameter demonstrated a dramatic degradation in the reconstruction performance: a single degree of freedom is apparently insufficient to encode the pulse profile efficiently.
With the aim of simplifying the following study, a~size of 2 is considered the best trade-off between maintaining sufficient fidelity in the bin-by-bin reconstruction of the waveforms while minimizing the latent representation~dimensionality.

The optimization of the VCAE parameters (i.e.,~weights and biases) is governed by a two-term loss function $\mathcal{L}$, which formalizes the trade-off between reconstruction fidelity and latent space regularization:
\begin{linenomath}
\begin{equation}
\mathcal{L} = \mathcal{L}_{\text{rec}} + \mathcal{L}_{\text{KL}}
\label{eq_loss}
\end{equation}
\end{linenomath}
The term $\mathcal{L}_{\text{rec}}$ is the MSE from Equation~(\ref{eq_MSE}), which enforces the network to reduce the bin-by-bin discrepancy between the input waveform $\textbf{x}$ and the output $\hat{\textbf{x}}$.
The term $\mathcal{L}_{\text{KL}}$ is the Kullback–Leibler (KL) divergence, which measures the statistical distance between the learned latent distributions and a prior standard normal distribution $\mathcal{N}(0, 1)$:
\begin{linenomath}
\begin{equation}
\mathcal{L}_{\text{KL}} = -\frac{1}{2} \sum_{j=1}^{k} \left( 1 + \log(\sigma_j^2) - \mu_j^2 - \sigma_j^2 \right)
\end{equation}
\end{linenomath}
where $\mu_j$ and $\sigma^2_j$ are the means and variances predicted by the encoder for the $j$-th latent component ($j=1,\dots,k$), with~$k$ being the latent space dimension.
The $\mathcal{L}_{\text{KL}}$ term acts as a structural regularizer which ensures a continuous and smooth latent embedding, preventing the encoder from collapsing into discrete points.
It is worth noting that no weighting factor is applied to balance the two loss~components.

A visual representation of the VCAE processing a synthetic waveform, generated as described in Section~\ref{subsec_toy}, is presented in Figure~\ref{fig_VCAE}. 

\begin{figure}[H]
\vspace{0.5cm}
\begin{adjustwidth}{-\extralength}{0cm}
\hfill
\includegraphics[width=18cm]{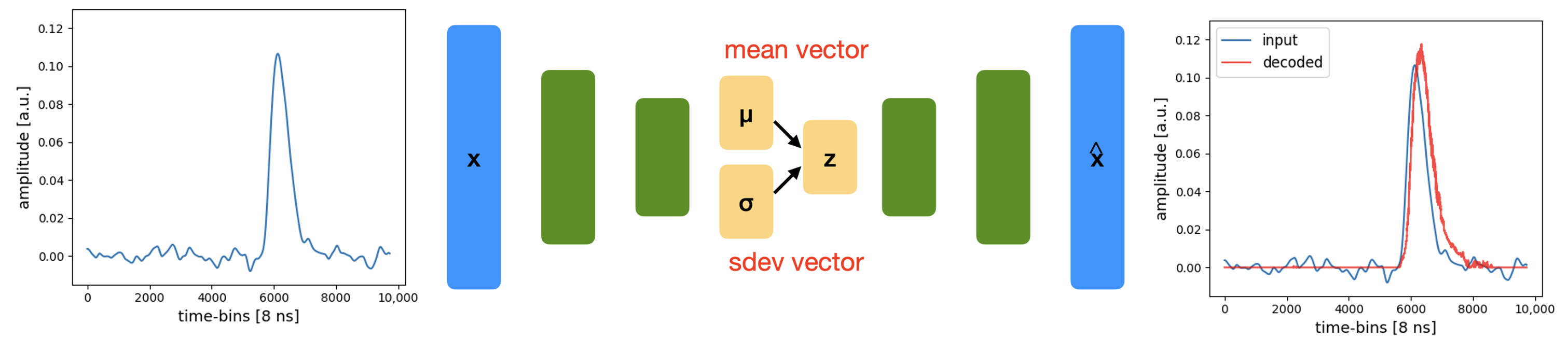}
\end{adjustwidth}
\caption{\label{fig_VCAE} A schematic view of the VCAE. On~the left, an example waveform is given as input; on the right, shown in red, the~reconstructed output.}
\end{figure}
\unskip 

\subsection{Cross-Validation and Training~Stability\label{subsec_train}}

For the training pipeline, a~total dataset of 15,000 synthetic waveforms was generated. The~dataset is divided into a training and validation set of $7500$ events, and~a testing set with the remaining $7500$ instances, reserved for the final performances~evaluation.

To ensure the statistical robustness of our results, a~5-fold cross-validation scheme is employed. The~training + validation dataset is partitioned into 5 disjointed folds and the complete training procedure is independently executed 5 times. In~each iteration, a~fold is reserved as the validation set to monitor the loss at the end of each epoch, while the remaining four folds constitute the training~set.

The VCAE architecture was found to exhibit a certain degree of sensitivity to the random initialization of the layers' weights and biases, leading to a failure of the learning procedure in approximately 20\% of the cases. In this context, successful training (also referred to as convergence of the model) is defined as the condition where the total training loss rapidly decreases over the first 10 to 20 epochs, confirming that the encoder--decoder pipeline has successfully started learning the data underlying distributions. Non-convergence is instead characterized by a total loss that either stagnates at a high initial value, oscillates without stabilizing, or~decreases only marginally over an extended number of epochs. 
To systematically mitigate this issue, a two-step training procedure has been~implemented:
\begin{enumerate}
    \item \textbf{Cooldown phase} (or pre-training): 10 VCAE instances are initialized with different random weights, drawn from the same probability distributions. Each model instance is trained for 20 epochs, monitoring the validation loss (that is, the loss calculated only on the validation set): the one achieving the lowest value is selected.
    \item \textbf{Training phase}: The selected ``best'' VCAE instance is further trained for 130 epochs. 
\end{enumerate}
The optimization is performed using the \texttt{ADAM} algorithm with an initial learning rate $\alpha = 10^{-3}$. In~addition, the~\texttt{ReduceLROnPlateau} callback is employed to scale down the learning rate by a factor of 0.5 if the validation loss fails to improve for 10 consecutive epochs. 
The result of the training process is shown in Figure~\ref{fig_losses}. 
The overall pipeline has been executed using the resources of the INFN Cloud infrastructure~\cite{url_INFNCloud}. 

\begin{figure}[H]
\includegraphics[width=0.9\textwidth]{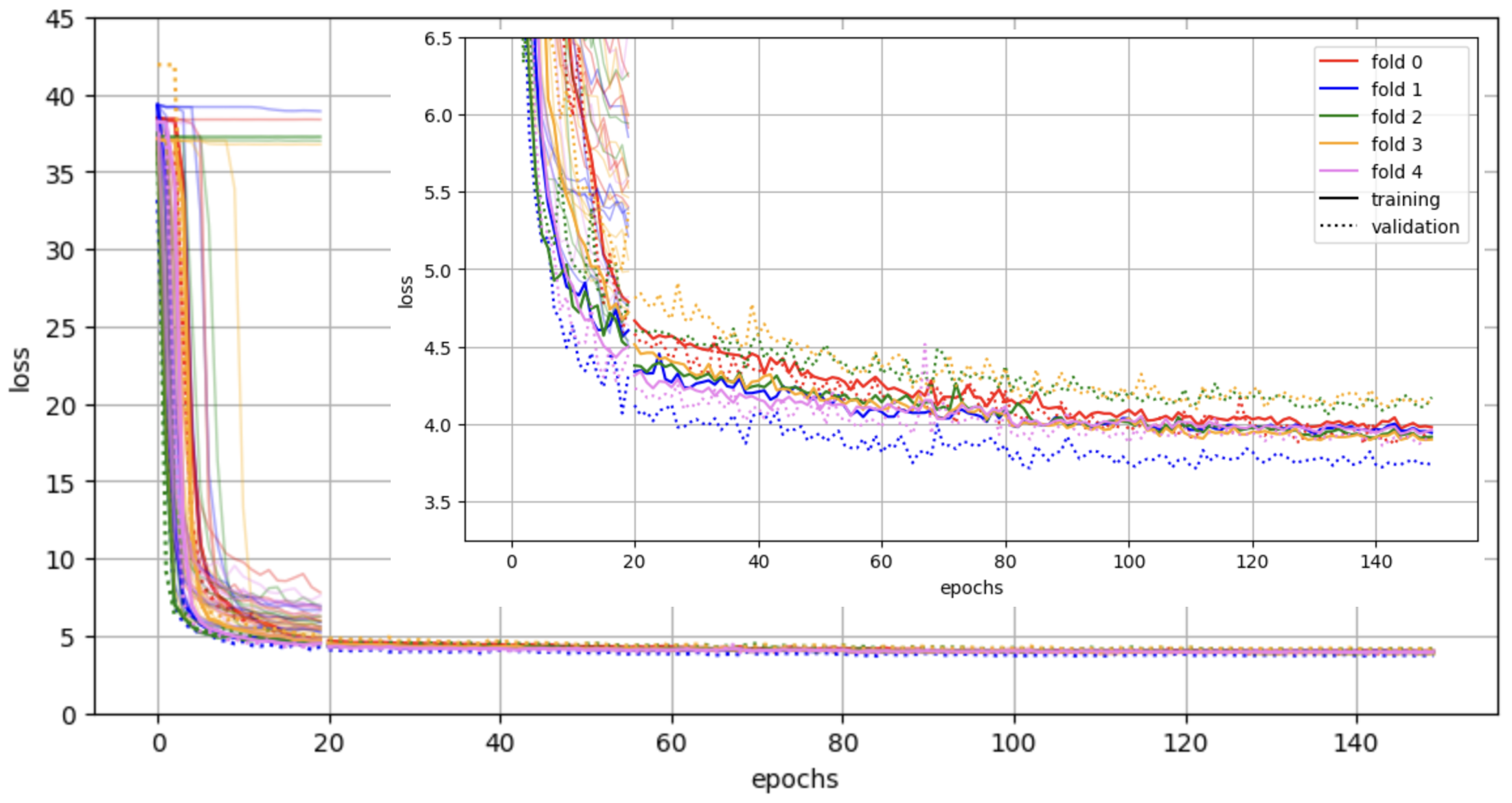}
\caption{\label{fig_losses} VCAE losses as a function of the training epochs, calculated according to Equation~(\ref{eq_loss}), for~the dataset of 7500 synthetic waveforms. The~5 folds results are shown with different colours, while the training and validation losses are represented with solid and dotted lines, respectively.
In the external plot and partially in the inset, it can be noticed that a fraction of the VCAE model instances does not converge during the cooldown phase; instead, the~training and validation losses decrease monotonically during the rest of the training, without~overfitting.
}
\end{figure}
\unskip


\section{Latent Space Representation and Pulse~Identification\label{sec_latent}}

After the training procedure described in Section~\ref{subsec_train}, we analysed the representation of the synthetic waveform dataset in the latent space. 
Throughout this section, quantitative results and distributions are explicitly displayed for the training process where the first fold was used as the validation set (referred to as Fold 0), since the latent space features were found to be fully consistent across the remaining four training~loops.

The encoder maps each input time-series into four statistical parameters: two latent means ($\mathbf{z}_{\text{mean}} = [z_{\text{mean}, 0},\ z_{\text{mean}, 1}]$) and two latent standard deviations ($\mathbf{z}_{\text{sigma}} = [z_{\text{sigma}, 0},\ z_{\text{sigma}, 1}]$), which characterize the Gaussian probability distributions from which the stochastic latent vector components $\mathbf{z} = [z_0,\ z_1]$ are sampled. Indeed, while the values of $\mathbf{z}$ used by the decoder fluctuate at each iteration, $\mathbf{z}_{\text{mean}}$ and $\mathbf{z}_{\text{sigma}}$ represent a (double) deterministic pair uniquely determined by the network's parameters for any given input waveform.
In this stage of our research, the~discrimination methodology was then developed focusing on the analysis of the $\mathbf{z}_{\text{mean}}$ vector.

As clearly visible in Figure~\ref{fig_latent_space_single}, the~marginalized distributions of $z_{\text{mean}, 0}$ and $z_{\text{mean}, 1}$ exhibit prominent accumulations within narrow (and distinct) intervals.  
Moreover, a dataset of noise-only time-series was built by subtracting the $S(t)$ term (see Equation~(\ref{eq_Wt})) from the waveforms included in the test set. Such noise-only waveforms are observed to be encoded exactly in these accumulation ranges by the trained VCAE network. Thus, we concluded that such localised peaks in the two distributions are related to the subset of waveforms where the physical signal is heavily buried in the non-Gaussian baseline noise.
As the signal amplitude increases, the~$z_{\text{mean}, 0}$ and $z_{\text{mean}, 1}$ values smoothly transition into extended tails, demonstrating that each latent variable independently encodes a sensitivity to the presence of genuine~pulses.

Furthermore, when the two $\mathbf{z}_{\text{mean}}$ parameters are projected onto a plane, their individual accumulations merge into a dense cloud, establishing a well-defined centroid at characteristic non-zero coordinates, which naturally corresponds to the localized region where noise-only waveforms are encoded, as~shown in Figure~\ref{latent_space_2D}.
High-amplitude signals radially diverge from this central cluster, clearly indicating that the VCAE mapped the signal intensity directly onto the geometry of the latent embedding: the smaller the physical pulse (using the Monte Carlo peak amplitude as a proxy), the~closer its representation lies to the noise-only~cluster. 

\begin{figure}[H]
\includegraphics[width=0.95\textwidth]{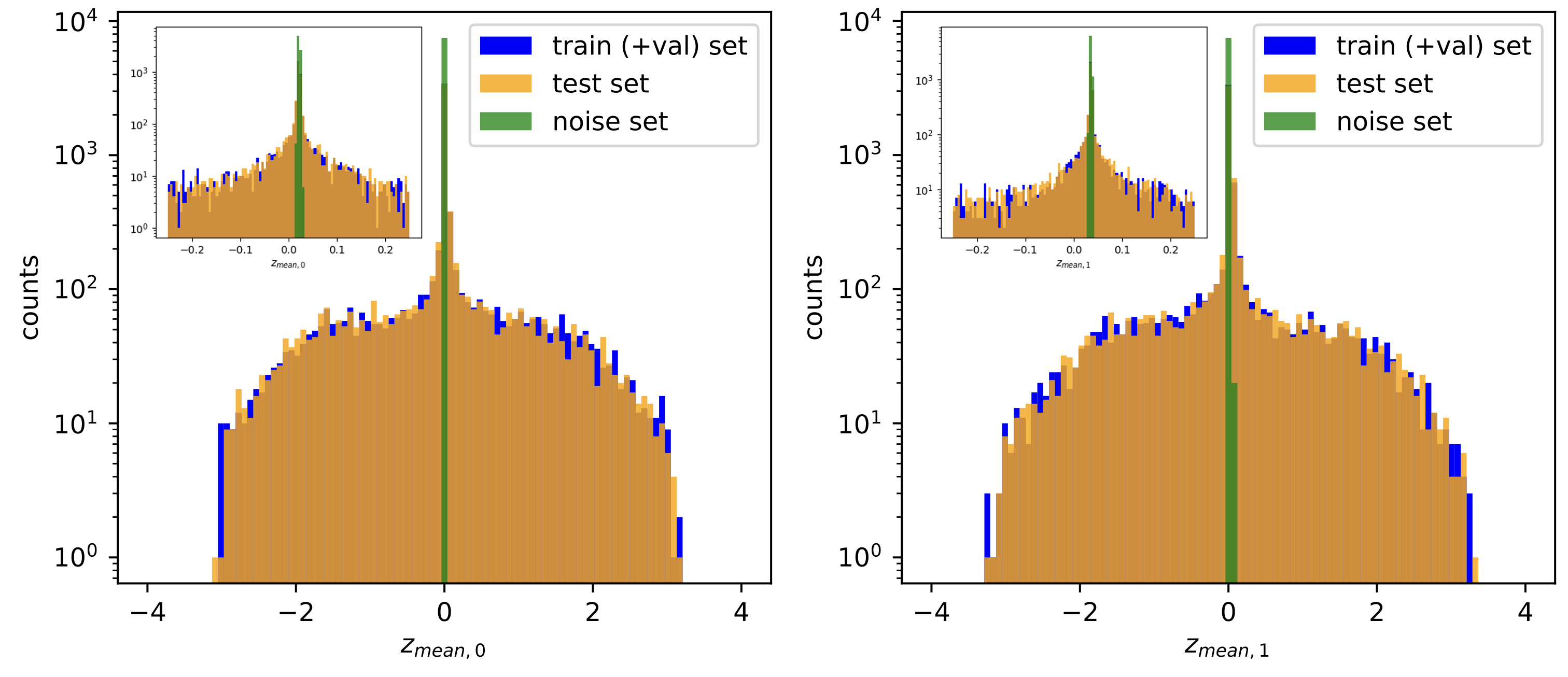}
\caption{\label{fig_latent_space_single} Distributions of the $z_{\text{mean}, 0}$ ({\textbf{left}}) and $z_{\text{mean}, 1}$ ({\textbf{right}}) latent space parameters using the Fold 0 model instance. In~blue, the~results for the training + validation set (7500 waveforms); in orange, for~the test set (7500 waveforms); in green, for~7500 noise-only waveforms (corresponding to the test set but devoid of signals). The~insets allow us to highlight the non-zero values of the accumulation regions, and~their correspondence with the encoding values of noise-only waveforms.
}
\end{figure}

Notably, the~encoding result reported in Figure~\ref{latent_space_2D} is obtained for the test set, confirming that the VCAE has successfully learned the intrinsic physical features of the waveforms rather than overfitting the training~events.

\begin{figure}[H]
\includegraphics[width=1\textwidth]{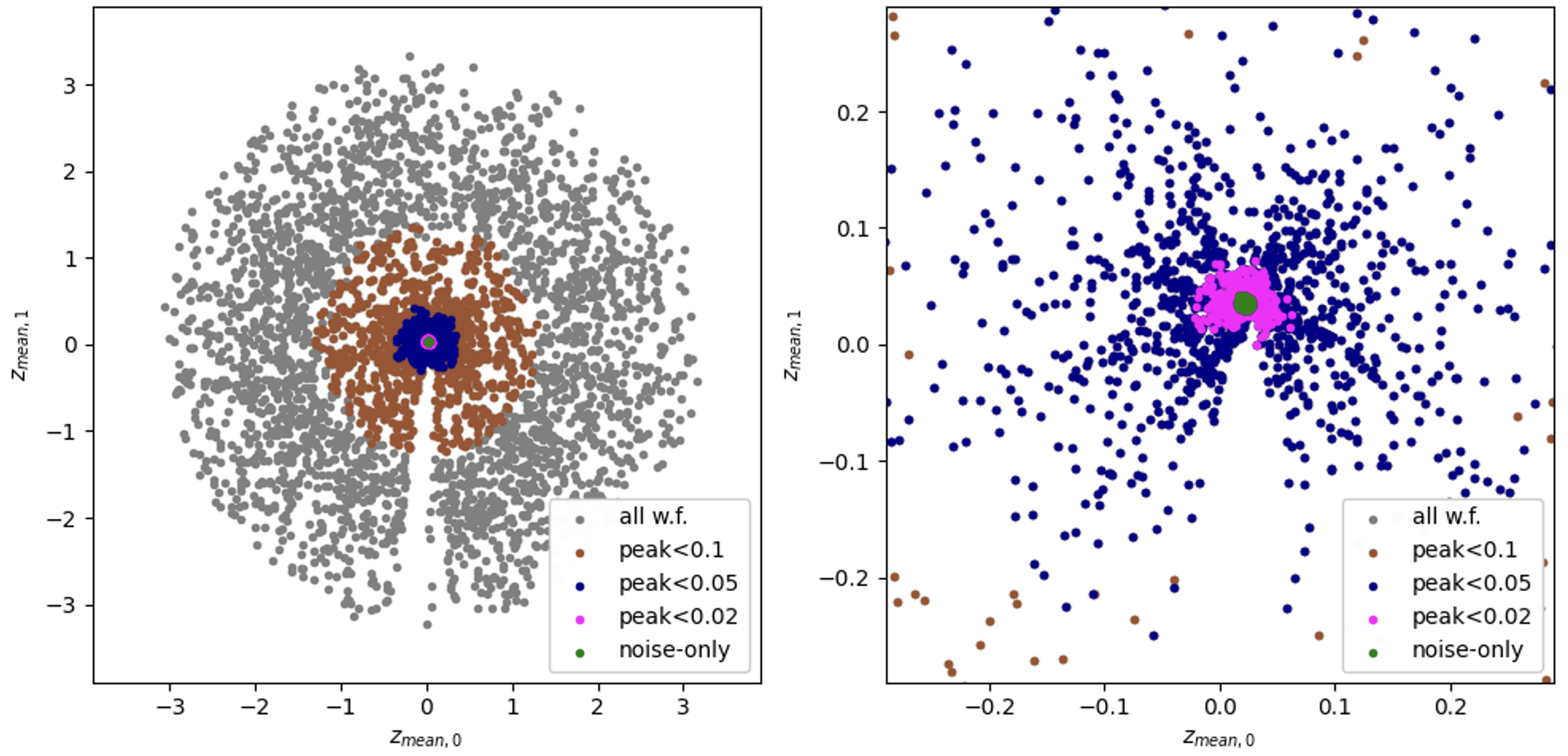}
\caption{\label{latent_space_2D} Two-dimensional latent space ($\mathbf{z}_{\text{mean}}$ plane) for Fold 0, where each point corresponds to the encoding of a synthetic waveform of the test set. The~{\textbf{left}} panel shows the overall distribution, while the {\textbf{right}} panel provides a zoomed view of the central region. 
Points are colour-coded based on their Monte Carlo peak (maximum signal amplitude).
Noise-only and small signal events are concentrated in the central accumulation, from~which higher-intensity signals radially diverge.
}
\end{figure}

\subsection{Distance-Based Selection~Criterion\label{subsec_criterion}}

To translate the observed qualitative behaviour into a rigorous classification methodology, we developed a distance-based selection algorithm. This data-driven procedure operates in two stages: an initial calibration phase, necessary to define the boundaries of the noise-only region in the latent space, followed by an operational phase where unlabelled waveforms are classified on an event-by-event~basis.

Firstly, a~``noise centroid'' $\mathbf{z}_{\text{noise}} = [z_{\text{noise}, 0},\ z_{\text{noise}, 1}]$ must be established. 
Rather than using the mean, which is susceptible to the tails and outliers, we used the mode of the distributions obtained encoding noise-only waveforms: $z_{\text{noise}, k} = \text{mode}(z_{\text{mean}, k}|{\text{noise-only}})$, calculated with 1000 bins in the range [$-$0.5, 0.5].

Once $\mathbf{z}_{\text{noise}}$ is fixed, the~separation of the encoded representation $\mathbf{z}_{\text{mean}}^{(i)}$ of any arbitrary waveform from the accumulation region can be evaluated by computing a standard Euclidean distance $d_i$ as:
\begin{linenomath}
\begin{equation}
d_i = \sqrt{\left(z_{\text{mean}, 0}^{(i)} - z_{\text{noise}, 0}\right)^2 + \left(z_{\text{mean}, 1}^{(i)} - z_{\text{noise}, 1}\right)^2}
\end{equation}
\end{linenomath}
To establish a distance threshold encompassing the noise-only region of the latent space, we evaluated the one-sided 99\% quantile of the distribution of distances for the noise-only set, denoting it as $d^*$, a~procedure reported in Figure~\ref{f8}a.
This threshold defines a circular boundary in the $z_{\text{mean}, 0}$ vs $z_{\text{mean}, 1}$ plane, centred at $\mathbf{z}_{\text{noise}}$ with radius $d^*$. 

Once the parameters $\mathbf{z}_{\text{noise}}$ and $d^*$ are determined, the~signal pulse identification for any unknown waveform proceeds via the following~steps:
\begin{enumerate}
    \item The waveform is processed by the trained VCAE encoder to extract $\textbf{z}_{\text{mean}}^{(i)}$;
    \item The distance $d_i$ from the noise centroid $\mathbf{z}_{\text{noise}}$ is calculated;
    \item The value of $d_i$ is evaluated against the threshold: if $d_i \le d^*$, the~waveform is tagged as noise-only; else, a~signal pulse is (considered to be) present.
\end{enumerate}

By construction, this procedure guarantees a 1\% False Positive Rate (FPR, i.e.,~the fraction of noise-only waveforms that are mis-labelled as signal candidates), limiting a priori the background~contamination. 
This holds only if the amplitude of the noise fluctuations $A_{noise}$ from Equation~(\ref{eq_Bt}) is kept constant while generating new waveforms. Indeed, the~described procedure, being data-driven, must be repeated when the parameters of the waveform generation are changed, in~order to recalculate $\mathbf{z}_{\text{noise}}$ and the $d^*$ value.

\vspace{-9pt}
\begin{figure}[H]
\begin{adjustwidth}{-\extralength}{0cm}
\centering
\hfill
\subfloat[\centering]{\label{fig_distance}\includegraphics[trim=0 0 0 30, clip, width=8.4cm]{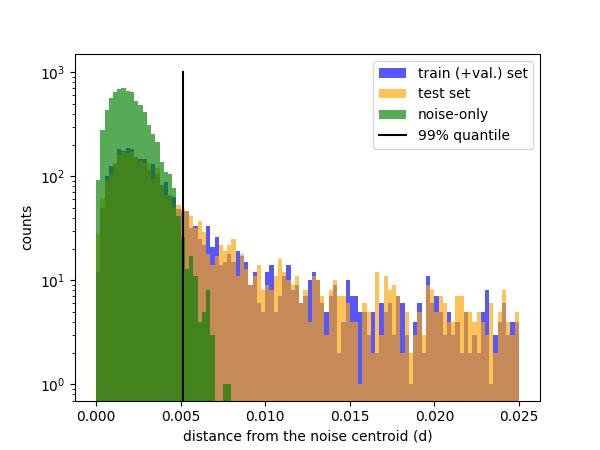}}
\subfloat[\centering]{\label{fig_eff0}\includegraphics[trim=0 0 0 30, clip, width=8.4cm]{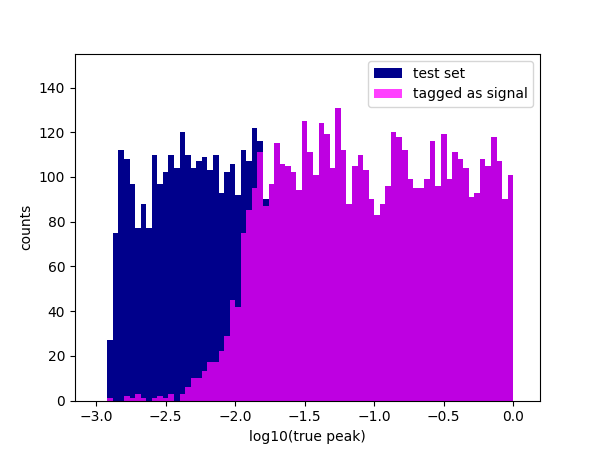}}
\end{adjustwidth}
\caption{(\textbf{a}) Distributions of the distance $d$ obtained using the VCAE trained with Fold 0 for validation. The~datasets employed are described in Figure~\ref{fig_latent_space_single}. 
The vertical line identifies the distance threshold $d^*$, corresponding to the 99\% quantile of the noise-only distribution.
(\textbf{b}) Distribution of the synthetic signal peak (maximum signal amplitude, indicated as ``\textit{true peak}'') for the waveforms of the test set. In~blue, the~overall histogram; in magenta, the~histogram obtained including only the values for waveforms correctly labelled as containing a signal by the VCAE-based method trained using Fold 0 for validation. Given that $A_{\text{S}}$ is distributed uniformly in the decimal logarithm (see Equation~(\ref{eq_signal_As})), the~dataset was represented using bins of equal width $\Delta \log_{10}(\mathrm{true~peak}) = 0.04$.
}\label{f8}
\end{figure}

Conversely, the~False Negative Rate (FNR, i.e.,~the fraction of waveforms containing a signal pulse that are mis-labelled as noise-only), and~by extension the selection efficiency $\varepsilon$ (calculated as $\varepsilon = 1 - \text{FNR}$), are inherently signal-dependent, as~low-amplitude pulses are mapped closer to the noise-only region and possibly within the $d^*$ boundary.
In Figure~\ref{f8}b, the~blue histogram represents the overall distribution of the generated pulses, while the magenta histogram denotes the subset of waveforms successfully tagged as containing a signal ($d > d^*$) by the VCAE-based selection criterion. 
In the higher-amplitude regime, the~selection efficiency asymptotically reaches 100\% (thus false negatives entirely disappear). As~the pulse amplitude shrinks toward the intrinsic noise fluctuations, fewer and fewer waveforms are recognised to contain a signal, as~expected.

To systematically evaluate our methodology (mainly in terms of the selection efficiency) and its stability, the~performances have been assessed across the 5-fold data validation scheme (detailed in Section~\ref{subsec_train}), as~reported in the next~section.

\subsection{Selection Efficiency Evaluation Across~Folds\label{subsec_eff_selection}}

The overall procedure described in the previous sections--including the study of the latent space representation of waveforms, the~evaluation of the noise centroid and the calculation of the threshold for the selection criterion--is independently performed on each of the 5 cross-validation folds.
Table~\ref{tab1} summarizes the quantitative parameters obtained across the folds: the coordinates of the noise-only centroid ($z_{\text{noise}, 0}$ and $z_{\text{noise}, 1}$) and the boundary distance $d^*$ corresponding to a 1\% FPR.
Moreover, the~minimum signal peak amplitude needed to achieve a selection efficiency $\varepsilon > 99\%$ is reported, the~calculation of which is explained in more detail later in this section.
Across folds, the~threshold values $d^*$ are of the same order of magnitude, and~the noise-only centroids are consistently located near the latent space origin, with~variations due to fluctuations in the training phases. 
To assess how these differences impact the signal tagging, the~efficiency $\varepsilon$ is evaluated on the test set using for each fold the corresponding trained model and selection~parameters.

\begin{table}[H] 
\caption{\label{tab1}Parameters and selection thresholds obtained from the latent space representations for the 5-fold cross-validation scheme. For~each fold, $z_{\text{noise}, 0}$ and $z_{\text{noise}, 1}$ denote the coordinates of the noise-only centroid in the 2D latent space, $d^*$ is the selection threshold distance corresponding to a 1\% False Positive Rate, and~the last column reports the true (synthetic) peak amplitude where the selection efficiency first exceeds 99\% (corresponding to the centre of a bin of width $\Delta \log_{10}(\mathrm{true~peak}) = 0.04$, as~for the binning procedure and efficiency calculation detailed in the text).}
\begin{tabularx}{\textwidth}{CCCCC}
\toprule
\textbf{Fold} & \textbf{$\boldsymbol{z_{\text{noise}, 0}}$} & \textbf{$\boldsymbol{z_{\text{noise}, 1}}$} & \textbf{$\boldsymbol{d^*}$} & \textbf{True Peak @ $\boldsymbol{\varepsilon > 99\%}$}\\
\midrule
fold\_0 & 0.022 & 0.036 & 0.0051 & 0.018\\
fold\_1 & 0.055 & 0.106 & 0.0192 & 0.022\\
fold\_2 & $-$0.035 & 0.003 & 0.0062 & 0.017\\
fold\_3 & $-$0.027 & 0.042 & 0.0070 & 0.046\\
fold\_4 & 0.014 & $-$0.020 & 0.0221 & 0.024\\ 
\bottomrule
\end{tabularx}

\end{table}

More specifically, the~$\varepsilon$ values are calculated directly as the ratio between the correctly tagged signal pulses and the total number of waveforms in the test set. In fact, the test dataset is completely independent of the (train + validation) 5-fold splitting, providing direct, unbiased performance comparison. Since the selection efficiency naturally varies as a function of the signal pulse amplitude, the~corresponding curves are computed as function of the true peak of the original (synthetic) signal, binning the test dataset logarithmically in the range [$-$3, 0] with bin-width $\Delta \log_{10}(\mathrm{true~peak}) = 0.04$ (as in Figure~\ref{f8}b), and~calculating the above-mentioned ratio within each~bin.

The resulting selection efficiency trends for all 5 folds are overlaid in Figure~\ref{fig_efficiencies}.
As expected, the~models show a consistent behaviour, characterized by a rapid transition from approximately zero efficiency into an asymptotic 100\% plateau as the signal peak increases.
Furthermore, four of the folds exhibit almost identical profiles, also reaching the efficiency plateau at similar peak amplitudes (as summarized in the last column of Table~\ref{tab1}). 
Only one fold displays a slightly slower turn-on and lower efficiency in the transition region, requiring a peak amplitude approximately doubled to reach $\varepsilon > 99\%$ with respect to the other folds.
This divergence is attributed to stochastic fluctuations during model training, which may alter the compactness of the noise-only region (and of the small-signals encoded representation) in the latent~space.

\begin{figure}[H]
\includegraphics[width=0.99\textwidth]{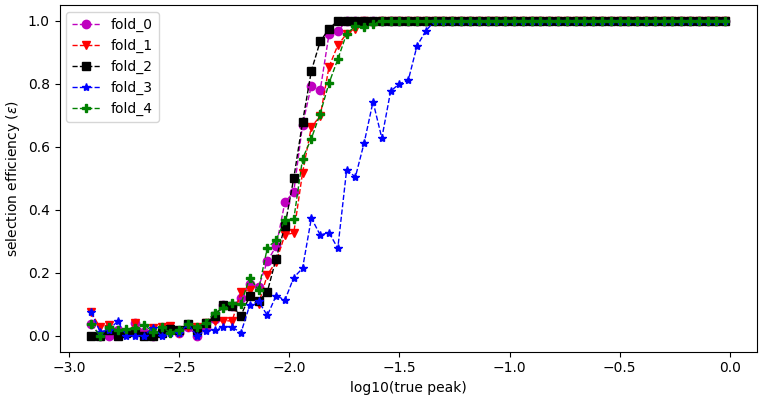}
\caption{\label{fig_efficiencies} Selection efficiency $\varepsilon$ as a function of the true (synthetic) signal peak amplitude, evaluated on the test set independently for all 5 cross-validation folds. The~$\varepsilon$ values are computed within logarithmic bins of width $\Delta \log_{10}(\text{true peak}) = 0.04$ between $-$3 and~0.}
\end{figure}
\unskip

\section{Results and~Discussion\label{sec_results}}

The performance evaluation, presented in the previous section, demonstrates that the VCAE-based methodology provides a highly effective pipeline for discriminating waveforms with signal pulses from background~noise. 

The final classification procedure is computationally light, requiring only model inference to obtain the encoded representation of the waveforms, since the VCAE has already been trained and the criterion calibrated. While keeping a fixed 1\% FPR, it achieves 100\% efficiency for waveforms with high-amplitude signals, confirming that the network successfully recognises the underlying pulse while suppressing noise fluctuations.
At lower signal amplitudes, the~selection efficiency exhibits a steep transition, dropping from the $\varepsilon=100\%$ plateau toward zero as the true peak of the synthetic signal approaches the magnitude of the noise fluctuations. 
However, this behaviour reflects an intrinsic physical limitation for any pulse-identification method: when the injected signal has an amplitude comparable to the baseline random pulses, it becomes fundamentally indistinguishable from pure noise. 
In the VCAE case, waveforms with progressively lower signal amplitudes are mapped increasingly closer to the noise-only cluster in the latent space, eventually falling within its~boundary.

The stability and reproducibility of the overall identification pipeline are confirmed by the consistency observed across the 5 cross-validation folds. 
The lower performance displayed by a single split (in terms of a slower transition of the selection efficiency and a higher peak amplitude needed to reach $\varepsilon > 99\%$) is attributable to the variability of the VCAE latent representation, which in this case affected the compactness and effective boundary of the noise-only cluster in the latent space. 
Nonetheless, the~response remains qualitatively uniform, demonstrating that the methodology is resilient to dataset partitioning and does not suffer from~over-fitting.

Compared to traditional techniques, the~primary advantage of the proposed selection methodology lies in its fully unsupervised, data-driven nature.
As outlined in Section~\ref{sec_exp}, conventional pulse-finding algorithms rely on thresholds on peak amplitude or integrated charge (with a preliminary deconvolution of the waveforms in ReD). 
While these methods perform more than adequately in high Signal-to-Noise-Ratio regimes, their efficiency degrades near the noise floor, where it becomes increasingly challenging to untangle stochastic baseline fluctuations from low-amplitude signals without incurring a high False Positive Rate (FPR).
Instead, the~VCAE training and latent space characterisation do not require a priori analytical models or empirical parameterizations of the expected signal shape, as~the network autonomously learns the underlying structures directly from the training data. 
In addition, the~waveforms are processed as a whole, ensuring high robustness against unexpected variations in pulse shape or detector noise levels. 
Naturally, in~the event of substantial changes in the detector operating conditions, a~re-training of the VCAE model would be~required.

\section{Summary, Conclusions, and~Outlook}\label{sec_conclusions}

In this work, we presented a signal-identification framework based on an unsupervised deep learning model, designed to discriminate waveforms containing a low-amplitude signal pulse from noise-only~baselines. 

The primary application envisioned for this methodology is the selection of low-energy (1--10~keV) nuclear recoil events in dual-phase argon TPCs, such as those employed in direct dark matter search experiments (Section~\ref{sec_exp}).
For low-mass dark matter candidates, the~prompt scintillation $S1$ signal falls below the detection threshold, leaving the delayed electroluminescence light pulse $S2$ as the only observable ($S2$-only events).
At increasingly lower energies, reliably identifying low-amplitude $S2$ pulses against electronic and photosensor noise still represents a challenge, especially in small-scale or non-ultrapure experimental~setups.

To establish a benchmark for our procedure, a~realistic dataset of synthetic waveforms was constructed (Section~\ref{subsec_toy}). These time-series were designed to reproduce the average of the calibrated readout channels of a TPC, combining high dimensionality ($\sim$10,000~time bins per waveform) with non-Gaussian baseline noise and injected signal pulses spanning a wide range of peak amplitudes.
An unsupervised Variational Convolutional Autoencoder (VCAE) architecture was then developed (Section~\ref{sec_VAE}) to process the raw waveforms. 
This network maps the input into a 2-dimensional latent space, autonomously learning the fundamental features of the training data and compressing them into an interpretable representation, even if not directly related to the physical signal parameters.
In this encoded representation (Section~\ref{sec_latent}), noise-only waveforms concentrate into a very localized region, and~by determining the centroid of this noise cluster, a~scalar metric based on the Euclidean distance in the latent plane was established. 
A threshold distance $d^*$, which encompasses the noise-only region and guarantees a 1\% False Positive Rate, was finally defined (Section~\ref{subsec_criterion}); consequently, waveforms encoded at distances greater than $d^*$ from the noise centroid are labelled as containing a signal.
The performance and stability of the complete methodology were evaluated using a 5-fold cross-validation scheme on the hold-out test set (Section~\ref{subsec_eff_selection}). 
The selection efficiency displays a 100\% plateau for signals well above the baseline noise, while it transitions smoothly toward zero for decreasing peak amplitudes.
Overall, these findings confirm that combining a VCAE latent representation with a distance-based criterion provides a reliable, data-driven framework for signal identification.
Eventually, to~determine the generalization capabilities of the developed methodology, a~systematic investigation under varying noise levels and pulse shapes is foreseen for future~work. 

Furthermore, while standard threshold-based methods (e.g., in~DS-50 and ReD) struggle near the level of noise fluctuations, the~proposed pipeline currently achieves robust selection efficiency on synthetic signals while maintaining a low false positive rate. Since our methodology is intrinsically data-driven, retraining the model directly on experimental data is expected to push the sensitivity down, toward the physical limits.
Building upon these observations, we plan to apply this algorithm to experimental datasets collected by the ReD dual-phase argon TPC. 
In the $S2$-only regime, low-energy nuclear and electronic recoils produce electroluminescence $S2$ pulses with indistinguishable shapes, as~the physical discrimination relies on the absent $S1$ signal and on time-of-flight measurements. Therefore, the network can be trained on a real dataset of $\gamma$-ray signals from the $^{252}$Cf source, enabling the VCAE to learn the actual features of measured waveforms. 
The noise-only region in the latent space will be calibrated using dedicated noise-only runs collected with the source shielded, and~the overall selection pipeline will be benchmarked on a high-purity neutron dataset. 
A comprehensive study of the selection efficiency, employing Monte Carlo simulated waveforms, is foreseen to rigorously determine the low-energy detection threshold.
Finally, upcoming developments will focus on extending the VCAE framework from using a single averaged waveform per event to simultaneously processing the waveforms from all 22 individual SiPMs composing the ReD optical readout. Capturing spatial-temporal correlations directly from raw multi-channel time series is expected to significantly enhance the signal identification capability, potentially reaching sensitivity down to $S2$ pulses produced by a single ionization~electron.

Beyond the specific applications to ReD, our data-driven methodology is intrinsically detector-agnostic and easily scalable. Consequently, it represents a versatile tool that can be exported to other dual-phase argon TPCs, offering a robust, template-free alternative for low-energy pulse selection in direct dark matter searches, such as those conducted within the DarkSide program, as~well as in broader noble-element detector~applications.

\vspace{6pt} 

\authorcontributions{
Conceptualization: G.A.A. and S.F.A.; 
methodology: G.A.A. and N.P.; 
data curation: S.M.P. and N.P.; 
software: G.A.A. and N.P.; 
writing---original draft preparation: G.A.A.; 
writing---review and editing: S.F.A., M.D.N. and S.M.P.; 
supervision: S.F.A. and A.R.T.; 
project administration: S.F.A. and A.R.T.;
funding acquisition: S.F.A. and A.R.T. All authors have read and agreed to the published version of the manuscript.}

\funding{This work is supported by ICSC -- Centro Nazionale di Ricerca in High Performance Computing, Big Data and Quantum Computing, funded by European Union -- NextGenerationEU (CUP E63C22001000006). 
The publication reflects the author’s view and neither the European Union or any Associated Partners are responsible for any use that may be made of the information contained therein.}

\institutionalreview{Not applicable.}

\dataavailability{
The source code implementing the model architecture, training pipeline, and~selection analysis described in this study is publicly available within the GitHub repository \url{https://github.com/Gialex05/DAIDREAM} (accessed on 20 August 2026), in~the \texttt{time\_series\_analysis/VAE\_application} directory (commit 1d12864) in the form of Python scripts and Jupyter notebooks. 
Furthermore, the~code required to generate the synthetic waveform datasets is provided in the form of Jupyter notebooks in the \texttt{time\_series\_analysis/synthetic\_waveforms} directory (commit 1d12864) of the same repository. The software versions employed are: Python 3.8.10, Numpy 1.24.3, SciPy 1.10.1, Keras-Tensorflow 2.13.1, JupyterLab 3.2.3 (kernel IPython 8.12.3).
} 

\acknowledgments{
The authors express their gratitude to the DarkSide Collaboration and, in~particular, to~the Recoil Directionality (ReD) working group for their valuable insights regarding the properties of experimental waveforms, which proved fundamental in developing our synthetic dataset, as~well as for their constructive feedback on the AE-based pulse identification~methodology.\\
During the preparation of this manuscript, the~authors used Google~Gemini in May 2026 (version 3.4 Flash and 3.1 Flash-Lite) and August 2026 (version 3.6 Flash) for polishing the text and streamlining the LaTeX typesetting process. The~authors have reviewed and edited the output and take full responsibility for the content of this publication.}

\conflictsofinterest{The authors declare no conflicts of interest.} 


\abbreviations{Abbreviations}{
The following abbreviations are used in this manuscript:
\\

\noindent 
\begin{tabular}{@{}ll}
DM & Dark Matter\\
WIMP & Weakly Interacting Massive Particle\\
LAr TPC & Liquid Argon Time Projection Chamber\\
DS-50 & DarkSide-50 (experiment)\\
DS-20k & DarkSide-20k (experiment)\\
ReD & Recoil Directionality (experiment)\\
SiPM & silicon photomultiplier\\
$S1$ & prompt scintillation signal \\
$S2$ & delayed electroluminescence signal\\
NR & nuclear recoil\\
ER & electronic recoil \\
PE & photoelectron \\
Sample & single time-bin of a recorded waveform (2~ns width for ReD data)\\
AE & autoencoder\\
MSE & Mean Squared Error\\
VAE & variational autoencoder\\
p.d.f. & probability density function\\
VCAE & variational convolutional autoencoder\\
KL & Kullback-Leibler divergence \\
Epoch & a full pass over the entire training set by the machine learning model under training\\
Distance & Euclidean distance in the 2-dim.~latent space, calculated from the noise-only centroid\\
FPR & False Positive Rate \\
FNR & False Negative Rate \\
\end{tabular}
}

\begin{adjustwidth}{-\extralength}{0cm}

\reftitle{References}


\PublishersNote{}

\end{adjustwidth}

\end{document}